\documentclass[bibyear]{aa}  
\usepackage{graphicx}
\usepackage{txfonts}
\newcommand{\be}{\begin{equation}}
\newcommand{\ee}{\end{equation}}
\newcommand{\bea}{\begin{eqnarray}}
\newcommand{\eea}{\end{eqnarray}}

\newcommand{\tet}{T_{e,\tau=\mu}}

\newcommand{\ang}{\,\AA}
\newcommand{\ha}{H$\alpha$}

\begin{document} 

\authorrunning{C. E. Alissandrakis etal}
\title{Center-to-limb observations of the Sun with ALMA
}
\subtitle{Implications for solar atmospheric models}
\author{C. E. Alissandrakis$^1$, S. Patsourakos$^1$, A. Nindos$^1$ \and T. S. Bastian$^2$
}

\institute{Department of Physics, University of Ioannina, GR-45110 Ioannina, 
Greece\\
\email{calissan@cc.uoi.gr}
\and
National Radio Astronomy Observatory (NRAO), 520 Edgemont Road, Charlottesville, VA 22903, USA
}

\date{Received ...; accepted ...}

 
  \abstract
 {}
{To derive information on the temperature structure of the solar chromosphere and compare it with existing models.}
{We measured the center-to-limb variation of the brightness temperature, $T_b$, from ALMA full-disk images at two frequencies and inverted the solution of the transfer equation to obtain the electron temperature, $T_e$ as a function of optical depth, $\tau$.} 
{The ALMA images are very similar to AIA images at 1600\ang. The brightness temperature at the center of the disk is 6180 and 7250\,K at 239 and 100\,GHz respectively, with dispersions of 100 and 170\,K. Plage regions stand out clearly in the 239/100 GHz intensity ratio, while faculae and filament lanes do not. The solar disk radius, reduced to 1\,AU, is $961.1\pm2.5$\arcsec\ and $964.1\pm4.5$\arcsec\ at 239 and 100\,GHz respectively. A slight but statistically significant limb brightening is observed at both frequencies.}
{The inversion of the center-to-limb curves shows that $T_e$ varies linearly with the logarithm of optical depth for $0.34<\tau_{100\,GHz}<12$, with a slope $d\ln T_e/d\tau=-608$\,K. Our results are 5\% lower than predicted by the average quiet sun model C of Fontenla et al. (1993), but do not confirm previous reports that the mm-$\lambda$ solar spectrum is better fitted with models of the cell interior.}

   \keywords{Sun: radio radiation -- Sun: quiet -- Sun: atmosphere -- Sun: chromosphere}

   \maketitle
%

\section{Introduction}

The solar atmosphere can be probed by measuring the specific intensity of the radiation either as a function of position from the center of the disk to the limb and/or as a function of frequency at the center of the solar disk.  Such observations allow, in principle, the inversion of the formal solution of the transfer equation and retrieve the source function of the radiation as a function of optical depth, which is the basis for the computation of empirical atmospheric models. 

An important difficulty in this procedure is the complex dependence of the absorption coefficient on the physical conditions and the departures from local thermodynamic equilibrium (LTE) in optical and UV wavelengths. The situation  is better in the radio range, which probes the solar atmosphere from the chromosphere to  the corona, thanks to the  Rayleigh-Jeans approximation to  the Planck function and the fact that the quiet Sun emission originates from thermal bremsstrahlung in LTE (see Shibasaki et al. 2011 and Wedemeyer et al.  2016 for details). In this context,  observations at millimeter wavelengths provide a valuable tool for atmospheric modeling. 

There is a long tradition of such observations with single-dish telescopes, but most of them suffered from low sensitivity and low spatial resolution; for typical dish sizes in the  range of 10-15 m the resulting resolution is of the order 1\arcmin\ at $\lambda=3$ mm (see Loukitcheva et al. 2004 and references therein). The resolution is improved either by observing at shorter wavelengths ({\it e.g.} observations at 0.85 mm reported by Bastian et al. 1993 and Lindsey et al. 1995 had a resolution of $\sim$20\arcsec) or by using larger telescopes ({\it e.g.} the Nobeyama 45-m telescope which yields a resolution of $\sim$20\arcsec\ at $\lambda=3$ mm; see Irimajiri et al. 1995). Furthermore, during solar eclipses, the edge of the Moon has been used to gain high resolution in one dimension ({\it e.g.} Belkora et al. 1992; Ewell et al. 1993). The reported brightness temperatures as a function of frequency show a large scatter due to inherent difficulties in the measurements and possible solar cycle effects.

According to empirical models the mm-$\lambda$ radiation comes mainly from the chromosphere. The model of Avrett and Loeser (2008) predicts that most of the emission at 3 cm and shorter wavelengths originates below the transition region (TR), with a small contribution from it and practically no contribution from  the corona. An important issue is that the observations are better fitted with models that represent the cell interior ({\it e.g.} Landi \& Chiuderi Drago 2003, Loukitcheva et al. 2004)  rather than with average models, which predict too much radio flux. Moreover, most millimeter-wavelength observations show less limb brightening than the homogeneous models predict (Shibasaki et al. 2011). Attempts to reconcile models with millimeter-wavelength observations invoke absorbing features such as chromospheric spicules ({\it e.g.}  Lantos and Kundu 1972; Selhorst et al. 2005). This approximation could be valid for the TR  and the corona where spicules are cooler than the ambient medium  but not necessarily so for the chromosphere ({\it e.g.} Tsiropoula et al. 2012).

\begin{figure*}
\sidecaption
\centering
\includegraphics[width=.7\hsize]{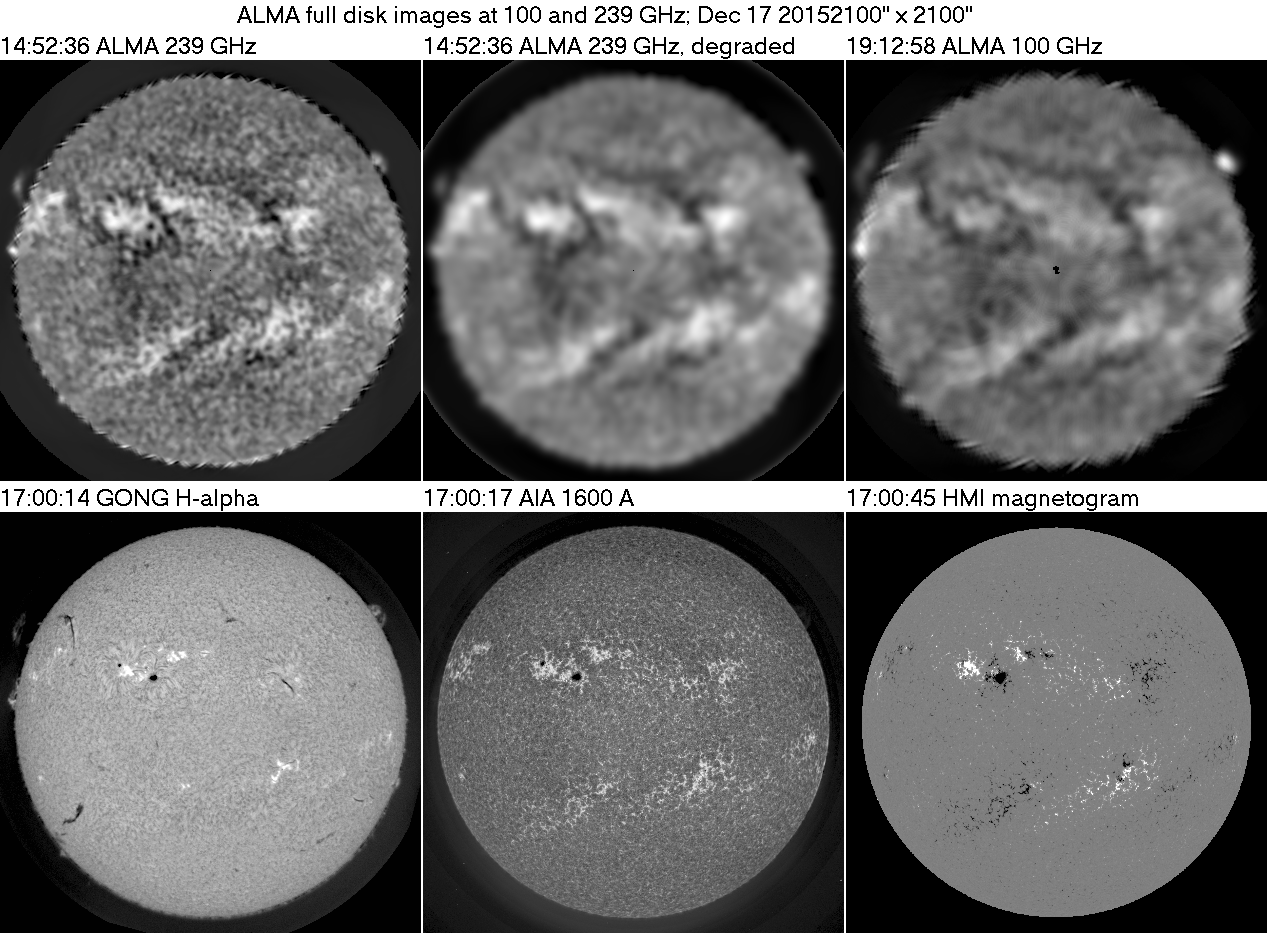}
\caption{ALMA full disk images on December 17, 2015, compared with images at other wavelengths. Top row, left to right: ALMA image at 239\,GHz; the same, degraded to the 100\,GHz resolution; ALMA image at 100\,GHz. Bottom row: \ha\ from the GONG network; 1600\ang\ image from AIA/SDO; magnetogram from HMI/SDO. The images have been partly corrected for center-to-limb variation. ALMA images are displayed with the same contrast.
}
\label{overview}
\end{figure*}

In this article we use recently released full-disk ALMA data at 100 and 239 GHz (3 and 1.26\,mm respectively) to study the center-to-limb variation of the emission and retrieve information on the variation of temperature with optical depth. Although these data have much lower  resolution than the interferometric ALMA data, the coverage of the entire solar disk, the low noise, the accurate absolute calibration and the great sensitivity of ALMA, make them important for our purpose. Moreover, we compare the mm-$\lambda$ disk structures with those in other wavelengths.

\section{Observations}

We analyzed data at 100 and 230 GHz obtained during solar commissioning activities on 2015 December 16, 17, 18, and 20 and released as ALMA Science Verification data in early 2017.  The observations were carried out in a  compact array configuration, which included twenty-two 12-m antennas and nine 7-m antennas. Single dish, fast-scanning observations from 12-m ``total power'' (TP) antennas covering the full solar disk were carried out simultaneously. The TP antennas observed the full solar disk (FD) within three minutes. The full with at half maximum (FWHM) of the single dish beam is about 27\arcsec\ at 239 and 60\arcsec\ at 100\,GHz, the sampling step 3 and 6\arcsec\ respectively and the field of view  2400\arcsec\ at both frequencies. The procedure for solar observations with ALMA is described in White et al. (2017).

The data set comprises of two FD images on December 16 and four on December 17 at 100 GHz, as well as one image per day on December 17, 18 and 20 at 239\,GHz. Fig. \ref{overview} shows ALMA full disk images at both frequencies for December 17, 2015. To facilitate the comparison between the two, we have added a 239 GHz image degraded to the 100 GHz resolution. We also give \ha\ and 1600\ang\ images as well as a magnetogram. The images have been partly corrected for center-to-limb variation, in order to enhance the visibility of the disk features and show prominences at the same time. We note that some irregularities appear at the limb of the ALMA images, which are due to the way that the antennas scan the sun. As the data close to the limb are not used here, these will not affect our results.

The four 100\,GHz images on December 17 gave us the opportunity to check the stability of the instrument. The total flux of the sun was remarkably stable, with an RMS of 0.2\%. On the disk, the RMS variation of the brightness temperature was not greater than 0.25\% of the average, while at the limb it was $\sim5$\% of the average, but this was mainly due to irregularities arising from the scanning procedure. The second image of December 16 was within 0.8\% of the December 17 images, however the first image of the same day was 3.3\% brighter and we preferred to ignore it in our analysis. Similarly, the three 239\,GHz images on December 18-20 were stable to within 0.3\%.

In the process of correcting the images by subtracting the center-to-limb variation we found that the solar disk was slightly elliptical in shape; this has nothing to do with any pole-equator asymmetry, since the major axis of the ellipse was inclined by about 45\degr\ east of the north pole. Although the effect was very small, amounting to a variation of the disk radius with amplitude of $\sim0.2$\%, it was clearly visible in the images corrected for center-to-limb variation and was taken into account. The origin of this effect, as well as of the irregularities near the limb probably lies in small pointing errors during the scanning process and/or in the algorithm of image reconstruction.

\begin{figure*}[ht]
\sidecaption
\centering
\includegraphics[width=.65\hsize]{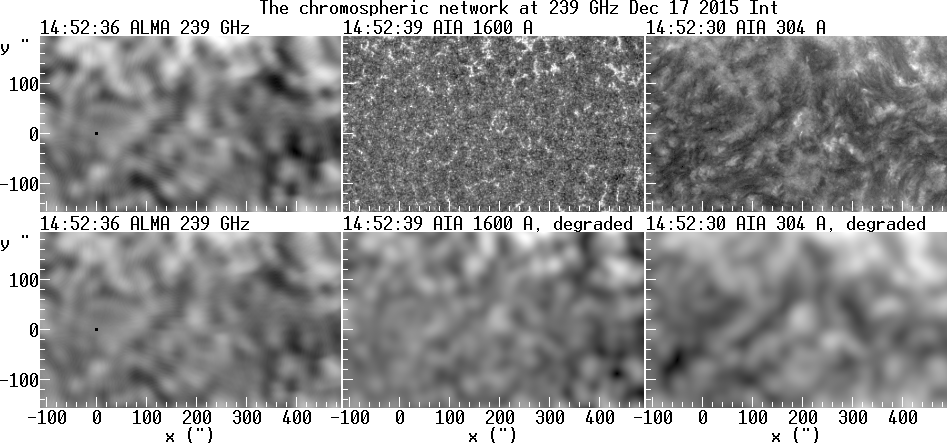}
\caption{The chromospheric network near disk center, from the ALMA FD image of December 17, 2015 at 239\,GHz, together with 1600\,\AA\ and 304\,\AA\ AIA images. In the bottom row the AIA images have been degraded to the ALMA resolution.
}
\label{network}
\end{figure*}
\begin{figure*}[!]
\sidecaption
\centering
\includegraphics[width=.65\hsize]{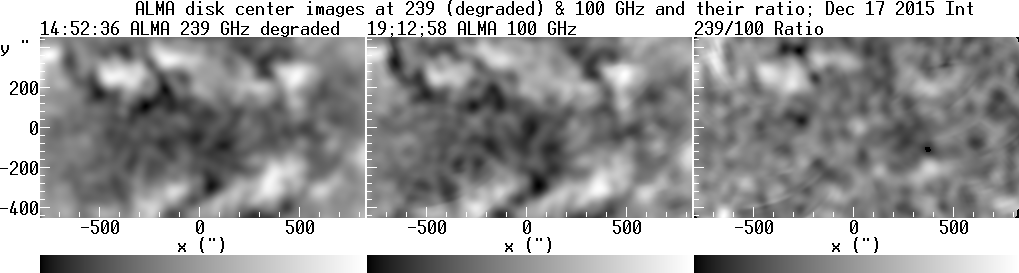}
\caption{The degraded 239 image, the 100 GHz image and their ratio. Values range from 6000 to 7380\,K, 6860 to 8380\,K and 0.83 to 0.93 respectively.
}
\label{ratio}
\end{figure*}

\section{Results}

The structure of the ALMA images  (Fig. \ref{overview}) is broadly consistent with previous works (Bastian et al. 1993; Iwai \& Shimojo 2015). Plage regions are the most prominent feature on the disk. The big sunspot in AR 12470, at N12W12, is well visible at 239 and barely discernible at 100\,GHz. Prominences are well visible beyond the limb, but large-scale neutral lines rather than filaments are seen on the disk as darker-than-average features. The chromospheric network is well visible at 239\,GHz and is very similar to the AIA 1600 and 304\,\AA\ images (Fig. \ref{network}), the correlation being slightly better with 1600\,\AA. This is a strong indication that the emission is formed at about the same level. We note that Bogod et al. (2015) found that cm-$\lambda$ structures correlate best with the 304\,\AA\ AIA band, which is formed higher,  in the low transition region. 

Fig. \ref{ratio} shows the 239/100 GHz intensity ratio; it ranges from 0.83 to 0.93, corresponding to a spectral index from $-0.21$ to $-0.083$. The average value of the ratio is 0.87, with a dispersion of 0.01. The ratio is higher for the active region plage, which has a flat spectrum, but not for the faculae in the south hemisphere. The dark lanes between the faculae are indistinguishable in the ratio image. We also note that the network appears more diffuse in the degraded 239\,GHz image than at 100\,GHz; since this is the opposite of what we expect because the 239\,GHz radiation is formed lower in the atmosphere, we suspect that the antenna beam at 100\,GHz is narrower than estimated.

\begin{figure}[t]
\centering
\includegraphics[width=\hsize]{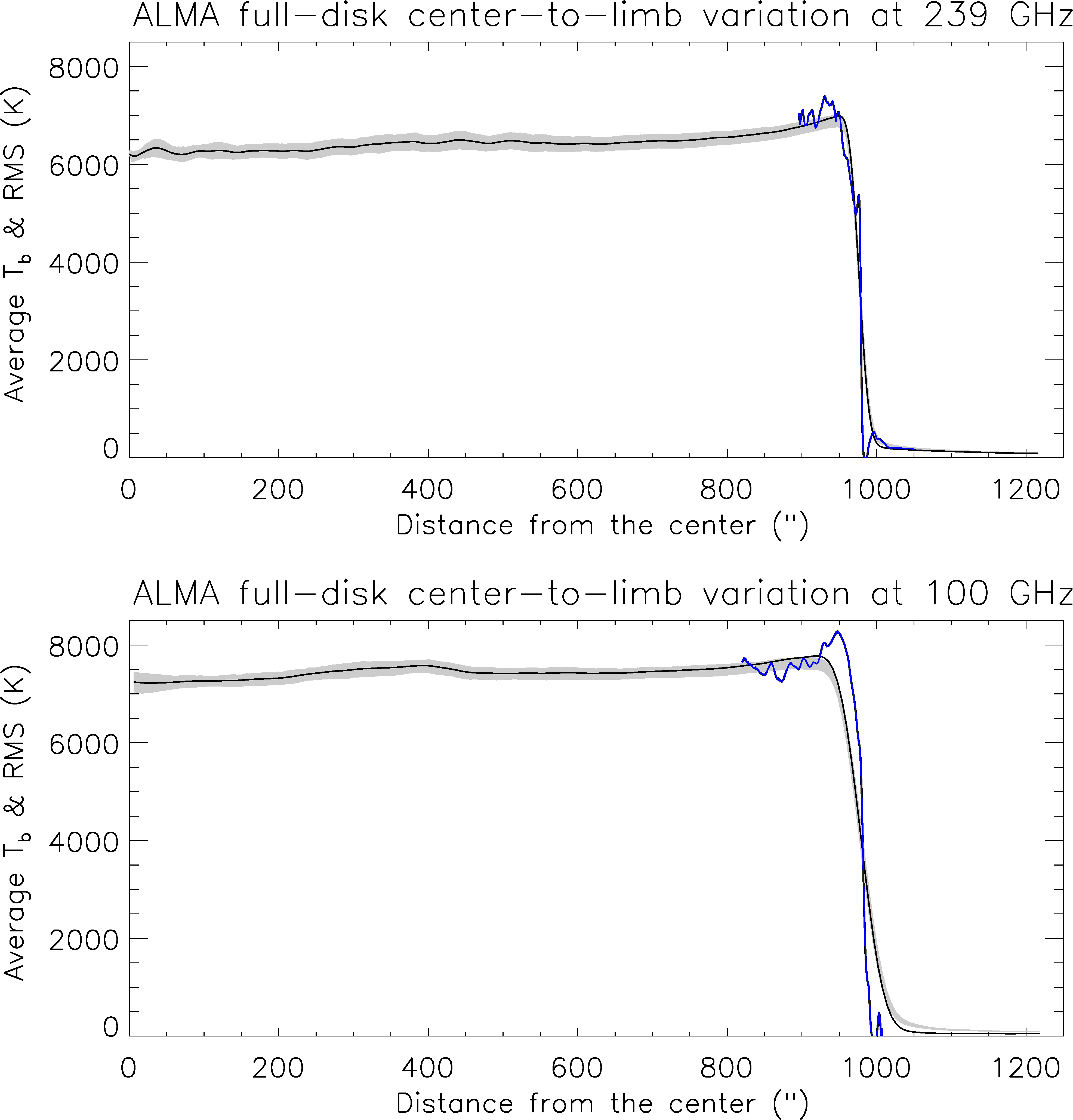}
\caption{Center-to-limb variation (CLV) for 239 and 100\,GHz. The gray band shows the measured values $\pm$ the corresponding RMS. The superposed blue curves near the limb are from high-resolution ALMA images. Curves in black show the CLV after correction for diffuse light.
}
\label{CLV}
\end{figure}

For the computation of the center-to-limb variation (CLV), we calculated the brightness temperature, $T_b$, as a function of distance from the disk center, $r$, after masking out the pixels that were either too bright (plage) or too dark (sunspot and filament channels); the results are shown in Fig. \ref{CLV}. The degrading effect of the antenna beam did not allow us to produce separate CLV curves for the network and cell interior, still a lower limit of their difference can be estimated from the intensity  RMS at each point, which also appears in Fig. \ref{CLV}.

In order to check how close to the limb our FD values are reliable, we computed CLV curves from two high resolution ALMA images that were included in the data set, one near the east limb on December 16 at 100\,GHz and one at the south pole on December 20 at 239\,GHz. We noticed that the flux of these images was higher than that of the FD, by 4\% for 239 and 15\% for 100\,GHz, apparently due to imperfect matching of the FD and high resolution flux. The high resolution CLV curves, corrected for this effect, are overplotted in Fig. \ref{CLV}. The comparison with the FD curves shows that we are safe in using measurements up to $r\simeq960$\arcsec, with a photospheric radius of 975.2\arcsec.

The CLV curves of Fig. \ref{CLV} show that $T_b$ does not drop to zero beyond the limb; at $r=1200$\arcsec\ (160$\times 10^3$\,km above the photospheric limb) it has a value of 100\,K at 239 and 93\,K at 100\,GHz. This effect could be due to (a) a coronal component, (b) the atmospheric background or, (c) extended wings in the antenna response (diffuse light). We can exclude the first possibility because the coronal contribution is very small; indeed, Alissandrakis et al. (2013), using differential emission measure derived from AIA, estimated the coronal contribution just beyond the limb to $\sim 8\,000$\,K at $\lambda=5.2$\,cm (see their fig. 9), which leads to $T_b\simeq25$\,K at 100\,GHz and 5\,K at 239\,GHz. The second possibility is rather unlikely, due to the careful ALMA calibration. We thus consider that the effect is due to the antenna response and thus it has a small influence on $T_b$ far from the limb. 

\begin{figure*}[ht]
\centering
\includegraphics[width=\hsize]{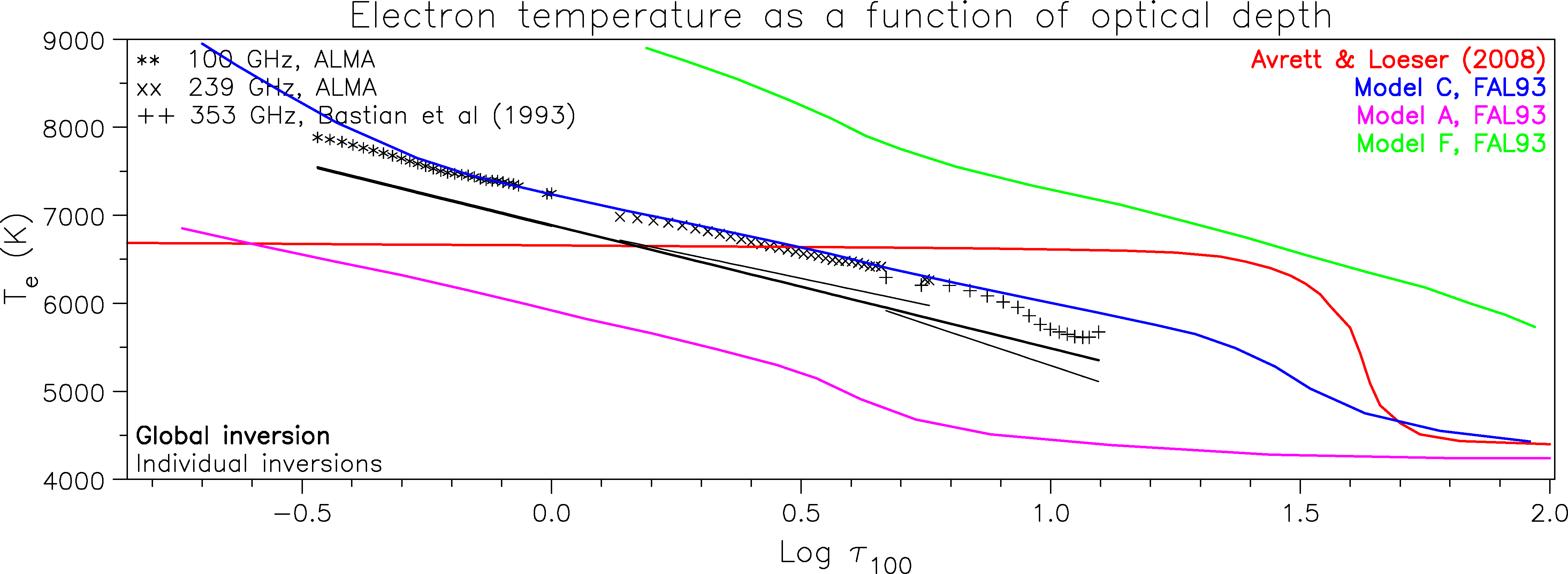}
\caption{The electron temperature as a function of the optical depth at 100\,GHz. The thick black line is the result of our inversion of the CLV curves of Fig. \ref{CLV}, together with the data of Bastian et al. (1993). Thin black lines are inversions of individual wavelength bands.The symbols mark the Eddington-Barbier solution of the transfer equation. The full lines in color were computed from the models of Fontenla et al. (1993, FAL93) and Avrett \& Loeser (2008).
}
\label{eddbar}
\end{figure*}

In order to improve the measurements closer to the limb, we performed a correction for diffuse light as follows: We first estimated the extended wings of the antenna response by computing the derivative of the CLV curve  near the limb; this was better fitted with a Voigt function rather than with a Gaussian and we attributed the extended wings of the Voigt function to the diffuse light and the Gaussian core to the combination of the antenna core and the actual limb profile. For the correction, we deconvolved with the Voigt function and reconvolved with the Gaussian core. The result is shown by the black curves in Fig. \ref{CLV}. We note that the correction did not change the slope of the CLV curve at the limb, while it slightly increased $T_b$ before the limb and reduced the diffuse light component beyond.

Both CLV curves are very flat, apart from some irregularities from remaining structures, with a slow but statistically significant increase towards the limb, as noted by  Bastian et al. (1993) at 353\,GHz. The average brightness temperature at a distance smaller than 30" from the center of the disk is 6180 and 7250\,K at 239 and 100\,GHz respectively, with dispersions of 100 and 170\,K. These fall within the ``cluster'' of values compiled by Loukitcheva et al. (2004) for this spectral range; we add that Iwai et al. (2017) measured $7700\pm310$\,K at 115\,GHz with the Nobeyama 45\,m telescope. The inflection point of the CLV curve is at $976.7\pm2.5$\arcsec\ and $979.8\pm4.5$\arcsec\ for 239 and 100\,GHz respectively, which is 1000 and 3300\,km above the limb; reduced to 1\,AU, the solar radio radius is 961.1 and 964.1\arcsec\ at 239 and 100\,GHz respectively. As expected, these values are smaller than the one reported by Selhorst et al. (2011) at 17\,GHz for the same phase of the solar cycle, which was $\sim975$\arcsec. 

\section{Inversion of the center-to-limb curve}
In principle, the variation of the electron temperature with optical depth, $T_e(\tau)$, can be obtained by inversion of the formal solution of the transfer equation for radio wavelengths:
\be
T_b(\mu)=\int_0^\infty T_e(\tau)\, e^{-\tau/\mu}\, d\frac{\tau}{\mu} \label{eqtran}
\ee
where $T_b$ is the brightness temperature, $\mu=\cos\theta$ and $\theta=\sin^{-1}( r/R_{\sun})$. 

In order to extend the range of $\mu$ and determine a single set of coefficients for all frequencies, we reduced all data to a common reference frequency, $f_{ref}$, using the fact that both the free-free and the H$^-$ absorption coefficients are proportional to $f^{-2}$. Hence a measurement at a frequency $f$ is re-mapped to:
\be
T_b\left(\mu(f/f_{ref})^2,f_{ref}\right)=T_b(\mu,f)
\ee
Moreover, we included the measurements of Bastian et al. (1993) at 353\,GHz (0.85\,mm) in our data set.

A first estimate of $T_e$ is provided by the Eddington-Barbier (E-B) relation, $T_e(\tau)\simeq T_{b,\mu=\tau}$. The result is shown in Fig. \ref{eddbar}, as a function of the logarithm of the optical depth at 100\,MHz. In the same figure we plotted $T_e(\tau)$ curves computed from standard atmospheric models, taking into account both the free-free (Zheleznyakov, 1970) and H$^-$ (Stallcop, 1974) absorption. 

Our E-B solution practically coincides with FAL93 Model C (average quit sun). However, E-B is just a first order approximation. For a more exact computation we preferred to assume a particular form for $T_e(\tau)$, which is a more stable procedure than a direct numerical inversion. For a logarithmic dependence of $T_e$ on $\tau$, suggested by both the E-B solution and the form of the model curves,
\be
T_e(\tau)=a_1+a_2\ln\tau	 \label{logform}
\ee
the brightness temperature is, from (\ref{eqtran}):
\be
T_b(\mu)=a_1+a_2(\ln\mu-\gamma)
\ee
where $\gamma$ is the Euler constant. The coefficients $a_1$ and $a_2$ can be determined by least square fit of the observations.

For the least square fit we used ALMA measurements up to $\theta\simeq70$\degr, in order to avoid regions affected by the beam near the limb and ignored small scale irregularities. The fit was very good, with an RMS of residuals of $\sim60$\,K, while the addition of a quadratic term in (\ref{logform}) did not change appreciably either the data fit or the inversion results, which shows that the solution is stable. Contrary to that, a power series expansion of $T_e(\tau)$ required at least six terms to give a good data fit and, moreover, it gave an oscillatory (hence unphysical) inversion result that did not converge even when ten terms were retained in the expansion. This justifies further our choice of logarithmic dependence of $T_e$ on $\tau$. 

Our inversion results are presented in Fig. \ref{eddbar}. In addition to the global inversion, we also performed inversions for individual frequency bands. The global inversion gave a slope of $a_2=-608$\,K. Inversions for individual wavelength bands are very similar to the global, with slopes of $-608$, $-519$ and $-818$\,K at 100, 239 and 343\,GHz respectively.

A first remark is that the inversion values are about 350\,K below the E-B solution, $T_{EB}$. This is because the linear expansion of $T_e$ used in this approximation is not accurate enough; adding a quadratic term with the second derivative, $\tet^{\prime\prime}$, to the Taylor expansion of $T_e(\tau)$ around $\tau=\mu$, we get
\be
T_e(\tau)\simeq\tet+(\tau-\mu) \tet^{\prime} +\frac{1}{2}(\tau-\mu)^2 \tet^{\prime\prime}
\ee
from which, applying (\ref{eqtran}), we obtain: 
\be
T_b(\mu)\simeq\tet + \frac{1}{2} \mu^2\tet^{\prime\prime} =\tet -  \frac{1}{2}a_2
\ee
where, in the last step, $\tet^{\prime\prime}$ was computed from (\ref{logform}). From this equation we obtain for the difference between the E-B solution and the 2nd order expansion:
\be
T_{EB} - \tet  \simeq  -\frac{1}{2} a_2  \label{BEdif}
\ee
 which evaluates to 304\,K for the value of $a_2$ given above and is very close to the difference between the E-B solution and our inversion.

Compared to the model predictions, our values are closer to the FAL93 model C (average quiet sun), $\sim5$\% lower; they are rather far from models A (cell interior) and F (network). The model of Avrett \& Loeser (2008) predicts too flat a chromosphere to be compatible with our measurements.

\section{Summary and conclusions}
Our study of the first ALMA full-disk images at 239 and 100 MHz showed that their morphology is very similar to that of AIA 1600\ang\ images, with plages, faculae and the chromospheric network being the most prominent features on the disk. We see prominences beyond the limb and dark lanes on the disk at the location of large scale neutral lines. Plage regions stand out clearly in the 239/100 GHz intensity ratio, while faculae and filament lanes do not. The brightness temperature at the center of the disk is 6180 and 7250\,K at 239 and 100\,GHz respectively, with dispersions of 100 and 170\,K. The solar disk radius, reduced to 1\,AU, is $961.1\pm2.5$ and $964.1\pm4.5$\arcsec.

Rather than comparing our center-to-limb measurements of the brightness with model predictions, we inverted them and obtained direct information on the variation of the electron temperature with the optical depth. Compared to spectral observations at discrete frequencies, CLV data have the advantage of covering a wider range of optical depths; the fact that they are obtained with the same instrument and at about the same time, gives them the additional advantage of homogeneity of observing conditions and phase in the solar cycle.

Our global inversion, combining the two ALMA frequencies and the data of Bastian et al. (1993) at 353 GHz, samples a $\tau_{100}$ range between 0.34 and 12. It is broadly consistent with a linear variation of electron temperature with the logarithm of the optical depth in the chromosphere, where the observed radiation is formed. Although we are $\sim5$\% below the model C of FAL93, we do not confirm previous results (see Shibasaki et al., 2011 and references therein) that the observations are better fitted with the cell interior model A. A potentially interesting aspect is that we do not reproduce the increased gradient of $T_b$ predicted by model C below $log(\tau_{100})=-0.3$ (Fig. \ref{eddbar}) and this may indicate that the temperature rise occurs higher than predicted by the model. However, our results in this $\tau$ range rely on 100\,GHz measurements  30 to 55\arcsec\ from the limb, which may have been affected by the 60\arcsec\ antenna beam.

It is difficult to establish a height scale on the basis of mm-$\lambda$ observations alone. Resorting to model C of FAL93, the height corresponding to our range of optical depth is from 980 to 1990\,km. The accurate determination of the position of the limb with high resolution ALMA observations will help to deduce empirical information on the height of formation of the mm-$\lambda$ emission. High resolution ALMA observations will also allow measurements of the ratio of network to cell intensity and feed this information to multi-component models, something that could not be done with the present, low resolution, full-disk data. Last but not least, as more ALMA frequency bands become available for solar observations, the range of heights sampled will extend from the upper photosphere to the low transition region, providing a unique tool for the study of solar atmospheric structure.

\begin{acknowledgements}
This paper makes use of the following ALMA data: ADS/JAO.ALMA\#2011.0.00020.SV.
ALMA is a partnership of ESO (representing its member states), NSF (USA) and
NINS (Japan), together with NRC (Canada) and NSC and ASIAA (Taiwan), and KASI
(Republic of Korea), in cooperation with the Republic of Chile. The Joint ALMA
Observatory is operated by ESO, AUI/NRAO and NAOJ. The authors gratefully acknowledge use of data from the AIA and HMI (SDO) as well as the GONG data bases.
\end{acknowledgements}

\end{document}